\documentclass[letterpaper, 10 pt, conference]{ieeeconf}  

\IEEEoverridecommandlockouts

\usepackage{cite}
\usepackage{amsmath,amssymb,amsfonts}
\usepackage{algorithmic}
\usepackage{graphicx}
\usepackage{textcomp}
\usepackage{xcolor}
\usepackage{mathtools}
\usepackage{subcaption}
\newcommand{\argmax}{\mathop{\rm arg\ max}\limits}

\title{\LARGE \bf
Bayesian Approach for Adaptive EMG Pattern Classification Via Semi-Supervised Sequential Learning
}

\author{Seitaro Yoneda$^{1}$ and Akira Furui$^{1}$
\thanks{This work was partially supported by JSPS KAKENHI Grant Number JP23H0343800.}%
\thanks{$^{1}$Seitaro Yoneda and Akira Furui are with Graduate School of Advanced Science and Engineering, Hiroshima University, Higashi-hiroshima, Japan
        (e-mail: seitaroyoneda@hiroshima-u.ac.jp; akirafurui@hiroshima-u.ac.jp).}%
}

\begin{document}

\maketitle
\thispagestyle{empty}
\pagestyle{empty}

\begin{abstract}
Intuitive human--machine interfaces may be developed using pattern classification to estimate executed human motions from electromyogram (EMG) signals generated during muscle contraction. 
The continual use of EMG-based interfaces gradually alters signal characteristics owing to electrode shift and muscle fatigue, leading to a gradual decline in classification accuracy. 
This paper proposes a Bayesian approach for adaptive EMG pattern classification using semi-supervised sequential learning. 
The proposed method uses a Bayesian classification model based on Gaussian distributions to predict the motion class and estimate its confidence. 
Pseudo-labels are subsequently assigned to data with high-prediction confidence, and the posterior distributions of the model are sequentially updated within the framework of Bayesian updating, thereby achieving adaptive motion recognition to alterations in signal characteristics over time. 
Experimental results on six healthy adults demonstrated that the proposed method can suppress the degradation of classification accuracy over time and outperforms conventional methods. 
These findings demonstrate the validity of the proposed approach and its applicability to practical EMG-based control systems.
\end{abstract}


\section{Introduction}
Surface electromyogram (EMG) signals are electrical activities that capture the action potentials of motor units in muscles from the skin surface, reflecting internal states of muscle contraction.
A classification model based on machine learning can estimate the correspondence between measured EMG signals and human's motion intention, aiding the construction of interfaces for operating devices such as myoelectric prostheses~\cite{tsoli2010robot,rezazadeh2012co,fougner2012control,irastorza2017emg}. 

The continuous utilization of EMG-based interfaces leads to a gradual decrease in motion recognition performance owing to variations in signal characteristics over time~\cite{sensinger2009adaptive,zhang2017robust}.
These variations are caused by distribution shifts between training and test data, which can be attributed to factors such as muscle fatigue~\cite{artemiadis2010emg}, change in motor imagery~\cite{samuel2017resolving}, and electrode shifts~\cite{betthauser2017limb,fougner2011resolving}.
Existing methods have predominantly employed static pattern classification models, which cannot adapt to such variations in signal characteristics.

To prevent the decline in classification performance over time, it is necessary to update the classification model sequentially based on changes in signal characteristics. 
For instance, Vidovic et al. and Zhu et al. demonstrated that updating the parameters of linear discriminant analysis (LDA) using a small amount of calibration data can sustain good performance over a prolonged period~\cite{vidovic2015improving,zhu2016cascaded}. 
In addition, Karrenbach et al. showed that a similar strategy is effective in transfer learning of deep neural networks~\cite{karrenbach2022deep}. 
However, these approaches require the periodic acquisition of labeled calibration data, a burden on users.

Various techniques have been proposed to update the classification models without using labeled calibration data within the framework of semi-supervised learning to achieve adaptive motion recognition without increasing the burden on users~\cite{zhang2013adaptation,okawa2022sequential,8692622,amsuss2013self,huang2017novel}. 
The typical approach is based on self-training, which predicts labels for test data using a classification model trained with labeled data, subsequently updating the model through retraining with the predicted labels as pseudo-labels~\cite{zhang2013adaptation,okawa2022sequential}. 
However, this method requires the constant retention of updating data for retraining, and the amount of stored data gradually increases unless appropriate data replacement is performed. 
In addition, since all pseudo-labels are used for updates, unreliable classification may lead to incorrect updating.
Therefore, if we could develop a method that enables the selective updating of the model for highly reliable data without retaining data instances, practical sequential learning for EMG would be realized.

This paper outlines a Bayesian approach for adaptive EMG pattern classification using semi-supervised sequential learning.
The proposed method employs a Gaussian classification model (GCM) to predict motion class labels and estimate their associated confidence.
The posterior distributions of GCM are then recursively updated via semi-supervised Bayesian sequential learning (SS-BSL) for test data with high predictive confidence, enabling adaptive motion recognition without the need for labeled calibration data.
Furthermore, information from previous trials is stored as the prior distribution parameters; thus, the model can be updated without retaining data instances.

\begin{figure}[t]
  \centering
    \includegraphics[width=\hsize]{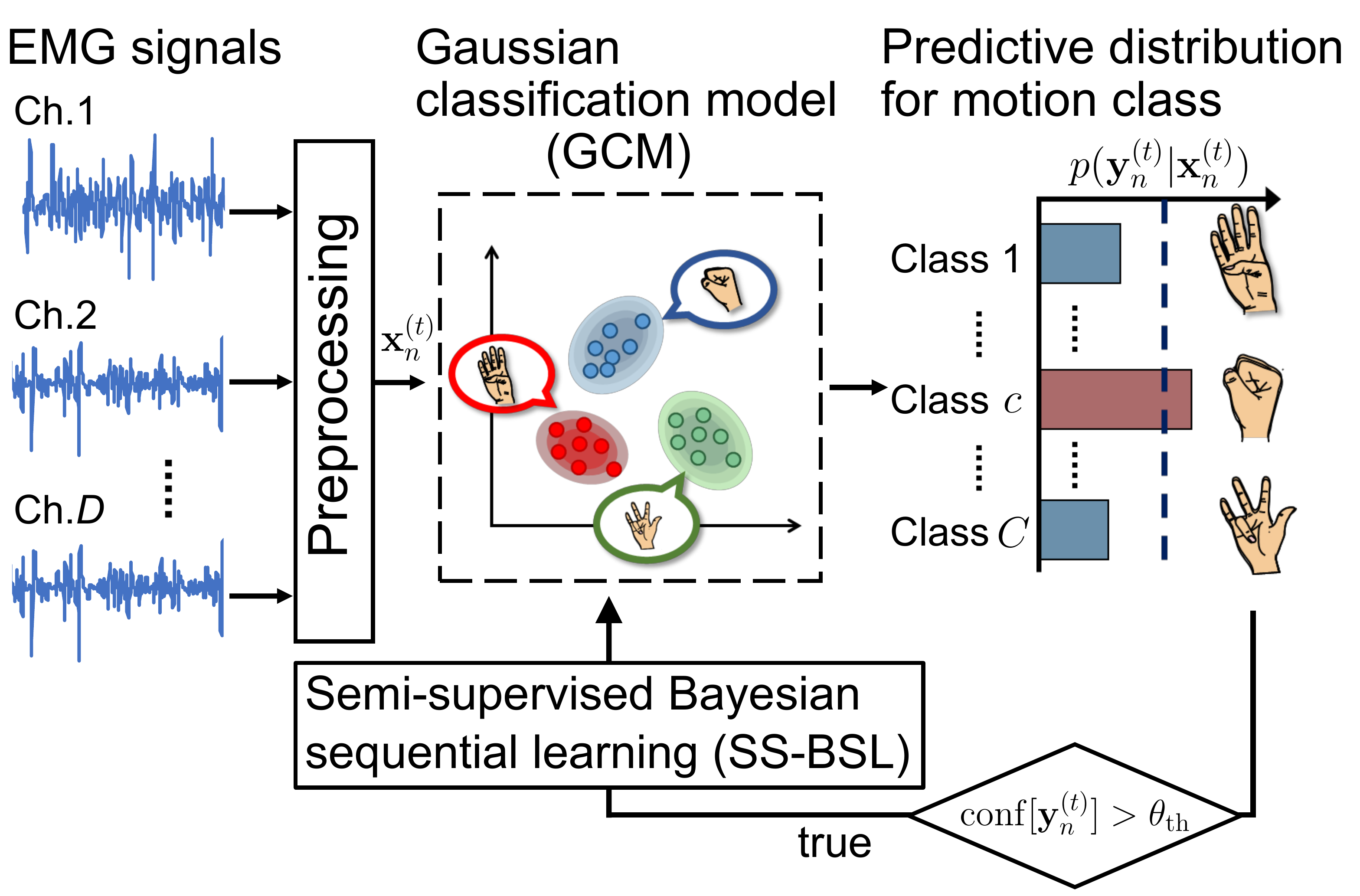}
  \caption{Overview of the proposed GCM with SS-SBL}
   \label{fig :overview}
\end{figure}

\begin{figure}[t]
    \centering
      \includegraphics[width=0.85\hsize]{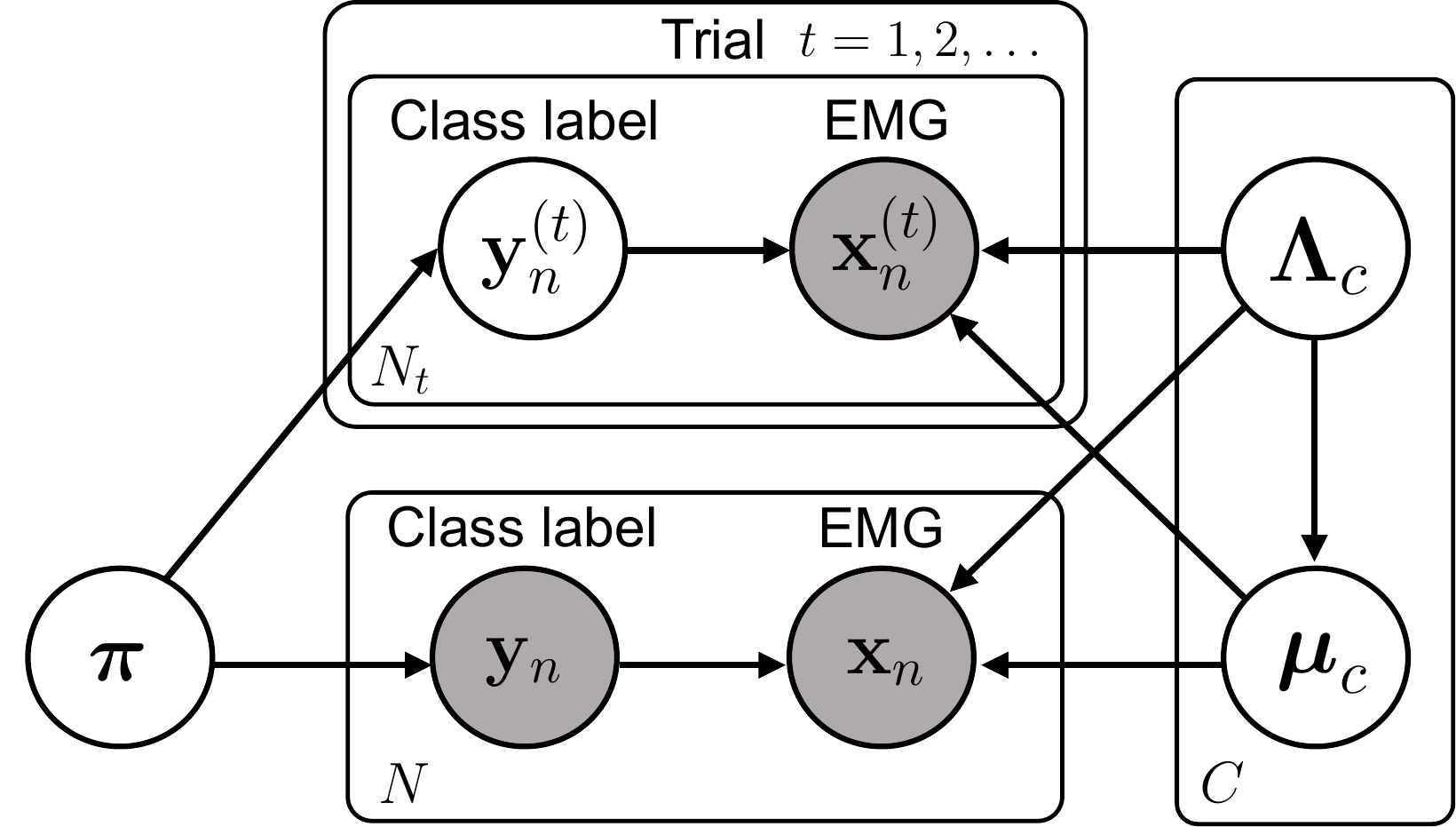}
    \caption{Graphical model of GCM for EMG classification. The gray and white nodes represent the observed and unobserved variables, respectively.}
     \label{fig :Graphicalmodel}
\end{figure}

\section{Proposed Classification Method}
Fig.~\ref{fig :overview} shows an overview of the proposed method.
The measured EMG signals at trial $t$ are first processed for feature extraction and then input into a trained GCM to compute class predictions and their associated confidence. 
The resulting output is then utilized to update the posterior distributions via SS-BSL. 
These procedures are repeated each trial, thereby adapting to changes in EMG signal characteristics over trial progress.

\subsection{Gaussian Classification Model (GCM)}
GCM describes the relationship between the EMG signal $\mathbf{x}_{n}\in\mathbb{R}^{D}$ captured at data point $n$ from $D$ electrodes and the corresponding motion class label $\mathbf{y}_n\in\{0,1\}^{C}$ ($C$ is the number of classes) using a Bayesian probabilistic model (Fig.~\ref{fig :Graphicalmodel}). 
The observed model of the EMG signal $\mathbf{x}_{n}$ for class $c \in \{1,\dots C\}$ is expressed via the following Gaussian distribution:
\begin{align}
    p(\mathbf{x}_n|\boldsymbol{\mu}_{c},\mathbf{\Lambda}_{c})=\mathcal{N}(\mathbf{x}_n|\boldsymbol{\mu}_{c},\mathbf{\Lambda}_{c}^{-1}), \label{Gaussian}
\end{align}
where $\boldsymbol{\mu}_{c}\in\mathbb{R}^{D}$ and $\boldsymbol{\Lambda}_{c}\in\mathbb{R}^{D\times D}$ represent the mean vector and precision matrix (the inverse of the covariance matrix) of each class, respectively.
To assign observations to each class, we introduce a one-hot representation for the motion class label $\mathbf{y}_{n}=\{y_{n,c}\}$. 
If $y_{n,c}=1$ for a given $c$, it means that the $c$-th class is selected.
Therefore, the distribution of $\mathbf{y}$ can be expressed via the following categorical distribution:
\begin{align}
    p(\mathbf{y}_{n}|\boldsymbol{\pi})=\text{Cat}(\mathbf{y}_{n}|\boldsymbol{\pi}) 
    =\prod_{c=1}^{C}{\pi}_{c}^{y_{n,c}}, \label{categorical}
\end{align}
where $\boldsymbol{\pi}=\{\pi_c\}$ is the mixing coefficients ($\pi_{c}\in[0,1]$ and $\sum_{c=1}^{C}\pi_{c}=1$).

To treat this model as a Bayesian model, a prior distribution is specified for each model parameter.
For ease of computation, we assign Gaussian-Wishart and Dirichlet distributions, which are conjugate priors for $\{\boldsymbol{\mu}_c, \boldsymbol{\Lambda}_c\}$, and $\boldsymbol{\pi}$, as follows:
\begin{align}
    p(\boldsymbol{\mu}_{c},\mathbf{\Lambda}_{c})&=\mathcal{NW}(\boldsymbol{\mu}_{c},\mathbf{\Lambda}_{c}|\mathbf{m},\beta,\nu,\mathbf{W}), \label{gauswhisert}\\
    p(\boldsymbol{\pi})&=\text{Dir}(\boldsymbol{\pi}|\boldsymbol{\alpha}), \label{dirikure} 
\end{align}
where $\mathbf{m}\in\mathbb{R}^{D}$, $\beta\in\mathbb{R}^{+}$, $\nu > D - 1$, $\mathbf{W}\in\mathbb{R}^{D\times D}$, and $\boldsymbol{\alpha} = \{\alpha_c\}$ are the hyperparameters and are common among classes.

\subsection{Semi-Supervised Bayesian Sequential Learning (SS-BSL) with Pseudo-labels}
In the proposed method, we introduce semi-supervised sequential learning to handle changes in the characteristics of EMG signals over time.
First, the initial learning of the model is conducted using a labeled training dataset.
Subsequently, the model is updated using sequential learning on newly acquired unlabeled test data. 
To achieve this, the predicted labels from the trained model are utilized as pseudo-labels.
The entire process, comprising both initial and sequential learning, is performed within a unified Bayesian updating framework.

\subsubsection{Initial learning}

Let us consider the initial learning of a model with a labeled training dataset $\mathcal{D}_0 = \{(\mathbf{x}_n, \mathbf{y}_n)\}^{N}_{n=1}$. 
This can be achieved by computing the posterior distributions of the model parameters, $\boldsymbol{\mu}$, $\mathbf{\Lambda}$, and $\boldsymbol{\pi}$, using $\mathcal{D}_0$. 
The posterior distributions for $\{\boldsymbol{\mu}_{c}, \boldsymbol{\Lambda}_{c}\}$ and $\boldsymbol{\pi}$ for each class can be calculated as follows:
\begin{align}
    p(\boldsymbol{\mu}_{c},\mathbf{\Lambda}_{c}|\mathcal{D}_0) &\propto p(\mathcal{D}_0 | \boldsymbol{\mu}_c, \boldsymbol{\Lambda}_c ) p (\boldsymbol{\mu}_c, \boldsymbol{\Lambda}_c),\\
        p(\boldsymbol{\pi}|\mathcal{D}_0) &\propto p( \mathcal{D}_0 | \boldsymbol{\pi}) p (\boldsymbol{\pi}).
\end{align}
From conjugacy, each posterior distribution is attributed to the following Gaussian-Wishart and Dirichlet distributions, which are of the same type as the priors:
\begin{align}
    p(\boldsymbol{\mu}_{c},\mathbf{\Lambda}_{c}|\mathcal{D}_{0})&=\mathcal{N}(\boldsymbol{\mu}_{c}|\hat{\mathbf{m}}_{c},(\hat{\beta}_{c}\mathbf{\Lambda}^{-1}_{c}))\mathcal{W}(\mathbf{\Lambda}_{c}|\hat{\mathbf{\nu}}_{c},\hat{\mathbf{W}}_{c}),\\    
    p(\boldsymbol{\pi}|\mathcal{D}_{0})&=\text{Dir}(\mathbf{\pi}|\hat{\boldsymbol{\alpha}}),
\end{align}
where $\hat{\beta}_c$, $\hat{\mathbf{m}}_c$, $\hat{\nu}_c$, $\hat{\mathbf{W}}_c$, and $\hat{\boldsymbol{\alpha}}$ are the hyperparameters of the posterior distributions and are defined as follows:
\begin{align}
    \hat{\beta}_{c}&=\sum_{n=1}^{N}y_{n,c}+\beta\label{fl-beta},\\ 
    \hat{\mathbf{m}}_{c}&=\frac{\sum_{n=1}^{N}y_{n,c}\mathbf{x}_{n}+\beta\mathbf{m}}{\hat{\beta}_{c}},\\
    \hat{\nu}_{c}&=\sum_{n=1}^{N}y_{n,c}+\nu,  \\    
    \hat{\mathbf{W}}_{c}^{-1}&=\sum_{n=1}^{N}y_{n,c}\mathbf{x}_{n}\mathbf{x}_{n}^{\top}+\beta \mathbf{m}\mathbf{m}^{\top}-\hat{\beta}_{c}\hat{\mathbf{m}}_{c}\hat{\mathbf{m}}_{c}^{\top}+\mathbf{W}^{-1},   \\
    \hat{\alpha}_{c}&=\sum_{n=1}^{N}y_{n,c}+\alpha. \label{fl-alpha}
\end{align}

\subsubsection{Sequential learning}
Let $\mathcal{D}_{t}=\{(\mathbf{x}^{(t)}_n, \tilde{\mathbf{y}}^{(t)}_n)\}_{n=1}^{N_{t}}$ denote the set of test data and pseudo-labels at trial $t \in \{1, 2, \ldots\}$. 
The pseudo-label $\tilde{\mathbf{y}}^{(t)}_{n}$ represents the predicted class labels determined by the trained model at trial $t-1$. The posterior distribution of each model parameter can be obtained by repetitively applying Bayes' theorem as follows:
\begin{align}
    p(\boldsymbol{\mu}_{c}, \boldsymbol{\Lambda}_{c} &| \mathcal{D}_{0}, \mathcal{D}_{1}, \ldots, \mathcal{D}_{t})\notag \\
    &\propto p(\mathcal{D}_{t} | \boldsymbol{\mu}_{c}, \boldsymbol{\Lambda}_{c})p(\boldsymbol{\mu}_{c}, \boldsymbol{\Lambda}_{c} | \mathcal{D}_{0}, \mathcal{D}_{1},\ldots, \mathcal{D}_{{t-1}}),\\
    p(\boldsymbol{\pi} &| \mathcal{D}_{0}, \mathcal{D}_{1},\ldots, \mathcal{D}_{t})\notag \\
    &\propto p(\mathcal{D}_{t} | \boldsymbol{\pi})p(\boldsymbol{\pi} | \mathcal{D}_{0}, \mathcal{D}_{1},\ldots, \mathcal{D}_{{t-1}}).
\end{align}
As in the initial learning, conjugacy results in the following form for each posterior distribution:
\begin{align}
    p(\boldsymbol{\mu}_{c}, \boldsymbol{\Lambda}_{c} |& \mathcal{D}_{0}, \mathcal{D}_{1},\ldots, \mathcal{D}_{t})\notag\\
     &= \mathcal{NW}(\boldsymbol{\mu}_{c}, \boldsymbol{\Lambda}_{c} | \hat{\mathbf{m}}^{(t)}_{c}, \hat{\beta}^{(t)}_{c}, \hat{\nu}^{(t)}_{c}, \hat{\mathbf{W}}^{(t)}_{c}),\\
     p(\boldsymbol{\pi} &| \mathcal{D}_{0}, \mathcal{D}_{1},\ldots, \mathcal{D}_{t}) = \mathrm{Dir}(\boldsymbol{\pi} | \hat{\boldsymbol{\alpha}}^{(t)}).
\end{align}
The hyperparameters of the posterior distributions at trial $t$ can be updated by substituting $y_{n,c}$ with $\tilde{y}^{(t)}_{n,c}$ and the hyperparameters of the prior distributions with those of the posterior distributions at trial $t-1$ in (\ref{fl-beta})--(\ref{fl-alpha}). 
Thus, the recursive substitution of the hyperparameters of the prior distributions can update the model with sequentially obtained test data for $t = 1, 2, \ldots$, resulting in adaptive motion recognition.

\subsubsection{Motion class prediction and assignment of pseudo-labels}
To perform classification on the test data $\mathbf{x}_{n}^{(t)}$ obtained at trial $t$, we can compute the following predictive distribution:
\begin{align}
  p(&\mathbf{y}_{n}^{(t)} | \mathbf{x}_{n}^{(t)}, \mathcal{D}_0, \mathcal{D}_{1}, \ldots, \mathcal{D}_{t-1})\notag \\ 
  &= \int p(\mathbf{y}_{n}^{(t)} | \mathbf{x}_{n}^{(t)}, \boldsymbol{\theta}_c) p(\boldsymbol{\theta}_c | \mathcal{D}_0, \mathcal{D}_1, \ldots, \mathcal{D}_{t-1}) \mathrm{d} \boldsymbol{\theta}_c, \label{kankeishiki}
\end{align}
where $\boldsymbol{\theta}_{c}=\{\boldsymbol{\mu}_{c}, \boldsymbol{\Lambda}_{c},\boldsymbol{\pi}_c\}$ represents the set of parameters, and $p(\boldsymbol{\theta}_c | \mathcal{D}_0, \mathcal{D}_1, \ldots, \mathcal{D}_{t-1})$ denotes its joint posterior distribution.
We determine the classified motion $\hat{c}_n$ as the class for which the class posterior probability, defined by the above predictive distribution, is maximized:
\begin{align}
  \hat{c}_n=\argmax_{c}p(y_{n,c}^{(t)}=1|\mathbf{x}_{n}^{(t)}, \mathcal{D}_0, \mathcal{D}_1, \ldots, \mathcal{D}_{t-1}).
\end{align}
In the proposed SS-BSL, the predicted label constructed based on $\hat{c}_n$ is used as the pseudo-label $\hat{\mathbf{y}}_n$. However, since the prediction results may contain errors, assigning all predicted labels as pseudo-labels may lead to inappropriate model updates.
To avoid this issue, we compute the confidence for the predictive distribution on each data point obtained at trial $t$ as follows:
\begin{align}
  \mathrm{conf}[\mathbf{y}_{n}^{(t)}] = \max_c p(y_{n,c}^{(t)} = 1 | \mathbf{x}_n^{(t)}, \mathcal{D}_0,  \mathcal{D}_1, \ldots, \mathcal{D}_{t-1}). \label{shikiichi}
\end{align}
We assign pseudo-labels for model updates only when $\mathrm{conf}[\mathbf{y}_{n}^{(t)}]>\theta_\mathrm{th}$, where $\theta_\mathrm{th}$ is an arbitrary threshold.
This approach excludes ambiguous data with a high likelihood of misclassification from the sequential learning.

\section{Simulation Experiments}
\subsection{Methods}
To assess the effectiveness of the proposed sequential learning approach in handling data with generating distributions that shift gradually, simulation experiments were conducted on artificial data comprising two classes.
Specifically, the data pertaining to classes $1$ and $2$ were randomly generated from two-dimensional Gaussian distributions denoted as $\mathcal{N}(\mathbf{x} | [-6+1.2t,3]^{\top}, 3.0 \mathbf{I})$ and $\mathcal{N}(\mathbf{x} | [6 - 1.2t,3]^{\top}, 3.0 \mathbf{I})$, respectively, where $\mathbf{I} \in \{0,1\}^{2}$ represents the identity matrix.
The mean vector of the generating distribution for each class varies as a function of the progression of trials, with $t= 1, 2, \ldots, T$.

The experiments generated a total of $300$ data points for each class at every trial $t$, thereby constructing the dataset $\mathcal{D}_{t}$. 
We set the upper limit of $t$ to $T=10$. 
First, we considered $\mathcal{D}_1$ as the training dataset and performed initial learning of GCM. 
Subsequently, each of the datasets $\mathcal{D}_2,\ldots,\mathcal{D}_{10}$ was treated as an individual test dataset and provided as input to the proposed method to perform class prediction and sequential learning with pseudo-labels.

We evaluated the classification accuracy of the proposed method (GCM with SS-BSL) for each test dataset. 
In addition, we also evaluated the classification accuracy when GCM was fixed after initial learning with no sequential learning performed.
The prior distribution parameters for GCM were set as $\mathbf{m}=\mathbf{0}$, $\beta=1$, and $\nu = D + 1$.
The covariance matrix based on the training dataset was set to $\mathbf{W}$, and $\boldsymbol{\alpha}$ was initialized with standard normal random numbers. 
The confidence threshold for sequential learning was set at $\theta_{\mathrm{th}} = 0.9$.

\begin{figure*}[t]
  \begin{tabular}{c}
    \begin{minipage}[t]{\hsize}
      \centering
      \includegraphics[keepaspectratio, scale=0.52]{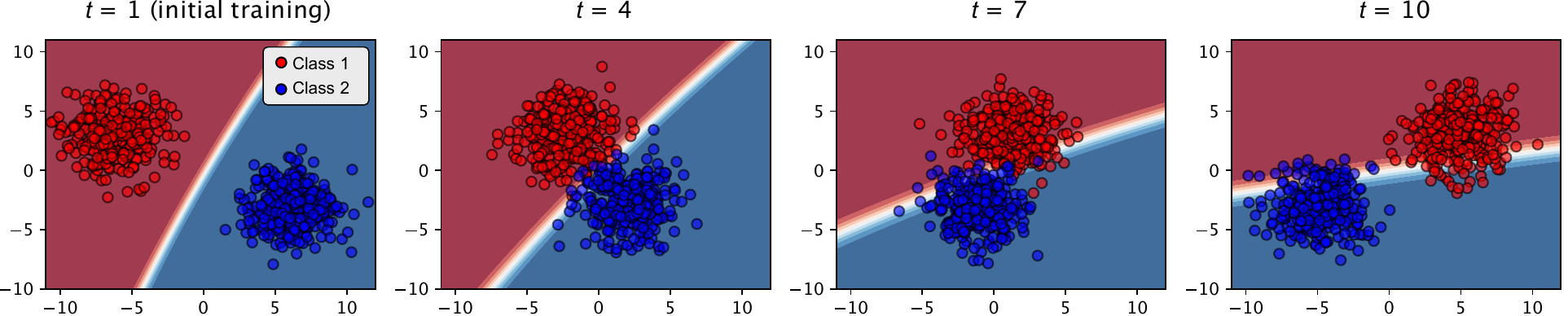}
      \centering{\subcaption{GCM with SS-BSL (\textbf{ours})}}
      \label{simulation_proposed}
    \end{minipage}\\\\
    \begin{minipage}[t]{\hsize}
      \centering
      \includegraphics[keepaspectratio, scale=0.52]{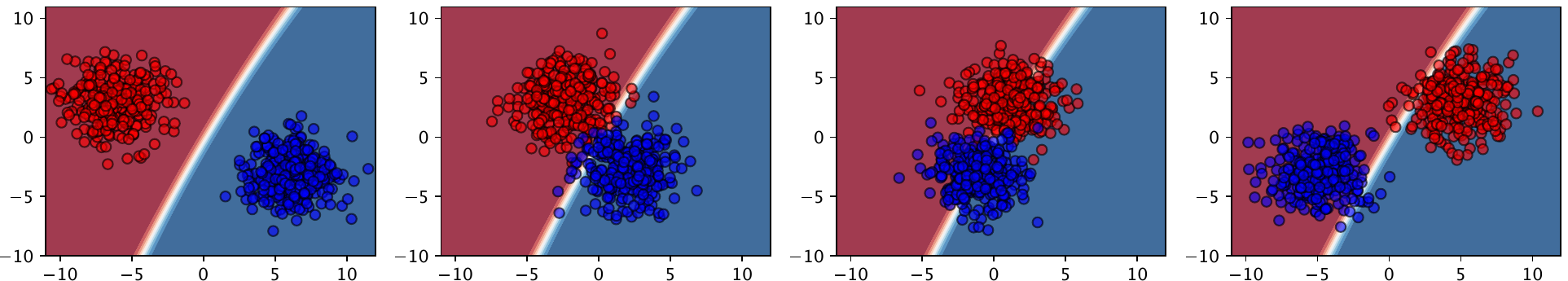}
      \centering{\subcaption{GCM}}
      \label{simulation_nonadaptation}
    \end{minipage}
  \end{tabular}
  \caption{Snapshots of the sequential learning process for each method. The red and blue areas indicate predictive
  distribution of class labels close to $1.0$ for classes $1$ and $2$, respectively.}
  \label{fig: simulation}
\end{figure*}

\subsection{Results and Discussion}
\begin{figure}[t]
    \centering
    \includegraphics[width=0.95\hsize]{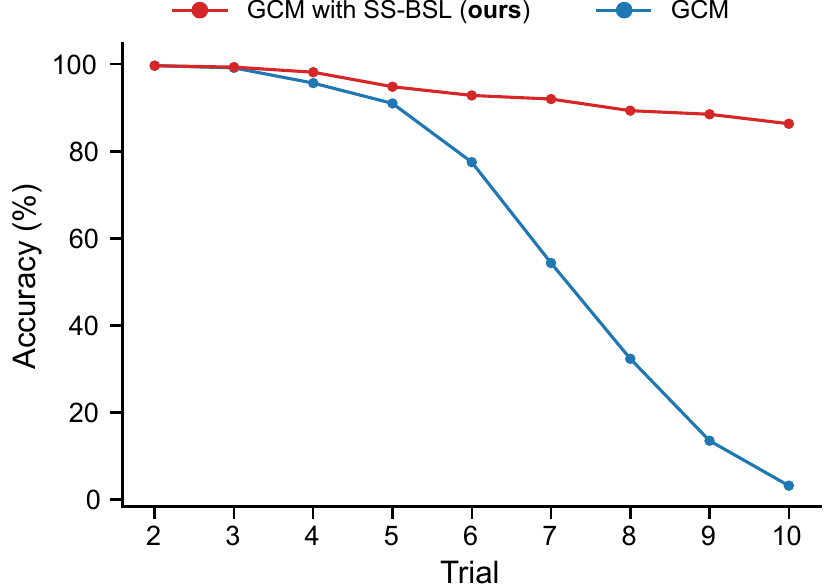}
      \caption{Per-trial classification accuracy for each method on simulated data}  
     \label{fig :simulation_accuracy}
\end{figure}

Fig.~\ref{fig: simulation} shows the scatter plot of the artificial data generated for each $t$, along with the predictive distributions of each class output by each method. 
The colored regions in the figure represent the predictive distributions, where the hues of red and blue indicate the probabilities of belonging to classes $1$ and $2$, respectively, being close to $1.0$. 
In addition, Fig.~\ref{fig :simulation_accuracy} shows the changes in the classification accuracy for each trial. 
The proposed method with sequential learning (GCM with SS-BSL) and the method without sequential learning (GCM) are represented by the red and blue lines, respectively.

In Fig.~\ref{fig: simulation}, the class $1$ data shifts toward the right, and the class $2$ data shifts toward the left as the trial progresses.
This indicates that the decision boundary derived from initial learning failed to accurately classify the data obtained from later trials.
Consequently, the accuracy of GCM without sequential learning noticeably decreased as the trial progressed (Fig.~\ref{fig :simulation_accuracy}).
In contrast, the proposed GCM with SS-BSL gradually adapts the decision boundary to the shifting data, resulting in relatively high accuracy even in the later part of the trial.
These results suggest that adaptive classification can be accomplished by sequentially updating the classification model using SS-BSL in response to the distribution shift of the test data.

\section{EMG Pattern Classification Experiments}
\subsection{Methods}
To verify the effectiveness of the proposed method for actual EMG signals, EMG pattern classification experiments were conducted. 
Six healthy adult males (mean age: $22.6 \pm 0.47$ years) participated in the experiments.
Four electrodes ($D = 4$) were attached at equal intervals around the arm near the elbow on the skin surface of the forearm. 
We used a wireless measurement system (Delsys, Trigno; sampling frequency: $2000$ Hz; bandwidth: $20$--$450$ Hz) to acquire and store EMG signals.
During the experiment, the participants were instructed to maintain a seated posture and perform $20$ trials of six different motions ($C = 6$): hand grasp, hand open, wrist flexion, wrist extension, supination, and pronation, with the right elbow resting on the desk. 
Each motion was measured for $7$ s, with rest periods of $5$ s between motions.

For feature extraction, full-wave rectification processing and smoothing using a second-order Butterworth low-pass filter with a cutoff frequency of $1$ Hz were performed.
To exclude the transition state from the resting posture, we removed the first $10\%$ of each data sample and used the remaining portion in the analysis. 
Of the $20$ trials, the first and second trials were treated as training trials, and the remaining trials were treated as test trials. 
In the analysis, all data from the training trials were used as the training dataset for initial learning of GCM.
Subsequently, the data from each test trial were input into the model, and sequential learning was conducted with the class label predictions and pseudo-labels. 
The parameters of the prior distribution were set the same as in the simulation experiment, and the confidence threshold for sequential learning was set to $\theta_{\mathrm{th}} = 0.9$.

We compared the classification accuracy of the proposed method (GCM with SS-BSL) with the following two baselines and two conventional methods.

\begin{itemize}
    \item \textbf{GCM} does not perform sequential learning, fixing the model after initial learning. We consider this baseline as the lower bound of performance for GCM-based methods.
    \item \textbf{GCM with fully supervised BSL (FS-BSL)} performs supervised sequential learning on the GCM using true class labels of test data instead of pseudo-labels. This method is unrealistic as the class labels of test data are unknown in practice. We consider this baseline as the upper bound of performance for GCM-based methods.
    
    \item \textbf{Adaptive LDA}~\cite{zhang2013adaptation} is a classification method that introduces semi-supervised sequential learning into LDA. Model updating employs a dataset combining training and test data. To suppress unexpected changes in parameters, a part of the updating dataset is fixed, and the rest is replaced cyclically. The replacement rate of the updating dataset must be set in advance, and we set it to $50\%$, which is recommended in~\cite{zhang2013adaptation}.
    
    \item \textbf{Self-training semi-supervised support vector machine (ST-S3VM)}~\cite{okawa2022sequential} is a classification method that combines self-training (ST) with the semi-supervised SVM (S3VM). Similar to adaptive LDA, this method combines the training dataset with a test dataset for updating the model. To prevent the expansion of the updating dataset with increasing trials, a maximum threshold was established to limit its size, and subsampling was executed on the entire dataset if the threshold was surpassed. In this experiment, the maximum threshold was identical to the sample size of $10$ trials. Three hyperparameters in S3VM were optimized by a $7 \times 7 \times 7$ grid search based on $2$-fold cross-validation on the training dataset. The search range was set to $C_1 = 2^{-13}, \ldots, 2^{-7}$ and $C_2 = 2^{-3}, \ldots, 2^{3}$ for the cost parameters for labeled and unlabeled data, respectively, and $\gamma = 2^{-20}, \ldots, 2^{-14}$ for the kernel coefficient for the RBF kernel.
\end{itemize}

  
\subsection{Results}

\begin{figure}[t]
    \centering
      \includegraphics[width=\hsize]{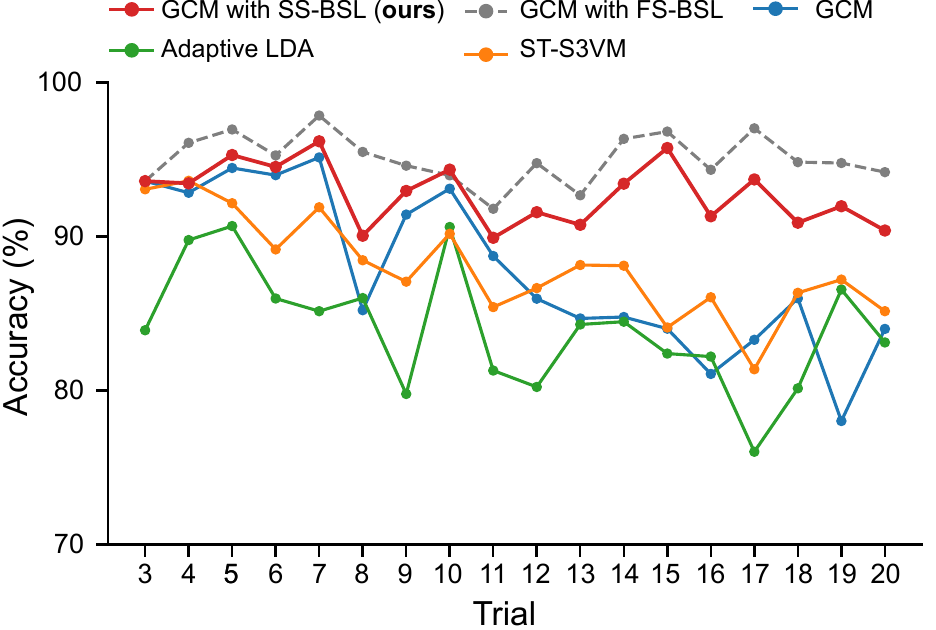}
    \caption{Averaged per-trial classification accuracy of each method over all participants}
    \label{fig :model_accuracy}
\end{figure}

\begin{figure}[t]
    \centering
      \includegraphics[width=\hsize]{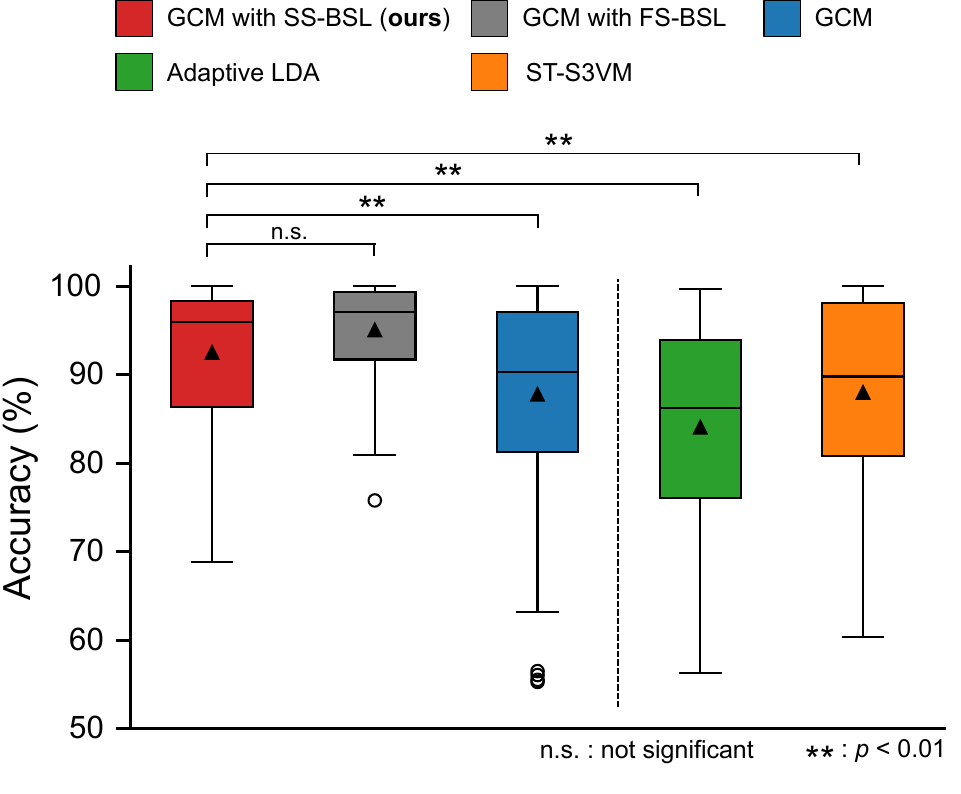}
    \caption{Boxplot of classification accuracy for each method over trials and participants}
    \label{fig: boxplot}
\end{figure}

Fig.~\ref{fig :model_accuracy} shows the changes in averaged classification accuracy for trial $t$. 
The results of GCM with FS-BSL, which is the upper-bound baseline, represent the ideal case where the true labels for test data, which should be unknown, are known. 
The proposed GCM with SS-BSL outperforms the simple GCM  without sequential learning and the conventional methods (i.e., adaptive LDA and ST-S3VM) in all trials. 

Fig.~\ref{fig: boxplot} summarizes the classification accuracy of each method in a boxplot over trials. 
The median and mean values for each boxplot are represented by a horizontal line and a triangle mark, respectively. 
The averaged accuracy for each method is as follows: GCM with SS-BSL: $93.4\%$, GCM with FS-BSL: $95.5\%$, GCM: $88.9\%$, adaptive LDA: $85.9\%$, and ST-S3VM: $87.9\%$. 
The figure also presents the results of pairwise comparison tests (significance level: $1\%$) with the proposed method as the control group. 
To account for individual differences, a linear mixed model was introduced where the classification method was set as a fixed effect, and the trials were treated as a random effect. 
The $p$-values between groups, calculated based on the estimated marginal means, were adjusted using the Holm method.

\subsection{Discussion}
In Fig.~\ref{fig :model_accuracy}, the accuracy of GCM without sequential learning decreases as the number of trials increases. 
This may be attributed to variations in the characteristics of EMG signals caused by muscle fatigue or changes in motor imagery, resulting in a gradually increasing mismatch between the distributions of training and test data. 
In contrast, the proposed GCM with SS-BSL suppressed the decrease in accuracy and maintained an accuracy of approximately $90\%$ in the later trials. 
These results suggest that the proposed BSL approach can adaptively recognize motions based on changes in EMG characteristics over time.

The classification accuracies of the conventional methods, adaptive LDA and ST-S3VM, were lower by $8\%$ and $6\%$, respectively, compared to the proposed method. 
These methods use all prediction results for model updates, which may decrease accuracy by incorporating uncertain data into the model. 
In particular, ST-S3VM determines support vectors by considering unlabeled data, causing significant changes in decision boundaries with the test dataset.
While this can be effective for long-term datasets with a large distribution shift of test data relative to training data~\cite{okawa2022sequential}, for short-term datasets, such as the one used in this experiment, excessive updating of the decision boundary may occur, causing a decrease in accuracy. 
Additionally, adaptive LDA sets the replacement rate for updating data to the value recommended in the previous study~\cite{zhang2013adaptation}, which may have been inappropriate for the dataset used in this experiment. 
In contrast to these conventional methods, the proposed GCM with SS-BSL can exclude uncertain data from the sequential learning process by determining the data used for model updates based on a confidence threshold. 
Furthermore, the BSL framework stores information from previous trials as prior distributions; thus, there is no need to additionally maintain updating data or adjust the replacement rate. 
These advantages enable the proposed method to achieve superior performance compared to the conventional methods in the sequential learning of EMG patterns.

Although the accuracy of the proposed SS-BSL is slightly lower than that of FS-BSL, which assumes the ideal scenario where the test labels are known, the discrepancy was not substantial (Fig.~\ref{fig :model_accuracy}). 
Moreover, the statistical test results did not reveal a significant difference between the two methods (Fig~\ref{fig: boxplot}). 
This indicates that the pseudo-labeling based on predictive confidence in the proposed method is effective. 
In practical applications, class labels for test data are unknown, and obtaining labeled calibration data separately from the test data can be challenging and increase the user's burden. 
Therefore, the proposed method, which does not require labeled data for updating, can maintain good performance over extended periods without causing any additional burden.

\section{Conclusion}
In this paper, we propose an adaptive EMG pattern classification method, GCM with SS-BSL, which utilizes Bayesian model-based sequential learning. 
The proposed method uses GCM to predict the motion class from EMG signals and estimate its confidence. 
Furthermore, pseudo-labels are assigned to data with high predictive confidence, and the posterior distributions of the model are sequentially updated by SS-BSL, thereby enabling adaptive motion recognition against changes in signal characteristics.

Simulation experiments demonstrated the applicability of the proposed method under conditions of gradual distribution shifts occurring in test data. 
EMG  classification experiments show that the proposed method outperforms conventional adaptive classification methods while suppressing accuracy degradation in continuous motion trials

In future research, we intend to increase the number of participants and trials and verify the effectiveness of the proposed method under more diverse conditions, such as changing postures during the motion execution.

\bibliographystyle{IEEEtran} 
\bibliography{reference} 

\end{document}